\documentclass[10pt]{article}
\usepackage{graphicx}
\usepackage{indentfirst,csquotes}

\usepackage{amssymb,amsthm,amsmath}
\usepackage{xcolor,paralist,hyperref,titlesec,fancyhdr,etoolbox}
\usepackage{subcaption}
\usepackage[normalem,normalbf]{ulem}
\usepackage{mathtools}
\usepackage[english]{babel}
\usepackage[T1]{fontenc}

\newcommand{\comment}[1]{}

\hypersetup{ colorlinks=true, linkcolor=black, filecolor=black, urlcolor=blue }

\begin{document}

\title{Kaon femtoscopy with L\'evy-stable sources from $\sqrt{s_{_{\rm NN}}} = 200$ GeV $\mathrm{Au}+\mathrm{Au}$ collisions at RHIC} %%%%%%%%%%%%
\author{Ayon Mukherjee\\\small{Department of Atomic Physics,}\\ \small{E\"otv\"os Lor\'and University (ELTE),}\\\small{Budapest, Hungary}}
\date{\today}
%\address{Department of Atomic Physics, E\"otv\"os Lor\'and University (ELTE), P\'azm\'any P\'eter stny. 1/A, H-1117, Budapest, Hungary}
%\email{ayon.mukherjee@ttk.elte.hu}

\maketitle

\let\thefootnote\relax
%\footnotetext{MSC2020: Primary 00A05, Secondary 00A66.} %%%%%%%%%%

\begin{abstract}
Femtoscopy has the capacity to probe the space-time geometry of the particle-emitting source in heavy-ion collisions. In particular, femtoscopy of like-sign kaon-pairs may shed light on the origin of non-Gaussianity of the spatial emission probability density. The momentum-correlations between like-sign kaon-pairs are, hence, measured in data recorded by the STAR experiment, from $\sqrt{s_{_{\rm NN}}} = 200$ GeV $\mathrm{Au}+\mathrm{Au}$ collisions at RHIC, BNL. Preliminary results hint at the, possible, existence of non-Gaussian, L\'evy-stable sources; and signal the, likely, presence of an anomalous diffusion process; for the identically-charged kaon-pairs so produced. More statistically significant studies, at lower centre-of-mass energies, may contribute to the search for the critical end point of QCD as well. 
\end{abstract} %%%%%%%%%

\bigskip

%%%%%%%%%%%%%%%%%%%%%%%%%%%%%%%%%%%%%%%%%%
\section{Introduction}
\label{s:intro}

%{\color{blue}
Following the discovery of the quark-gluon plasma, one of the main thrusts of high-energy nuclear-physics has been the understanding and exploration of the space-time geometry of the particle-emitting source created in heavy-ion collisions~\cite{Lisa:2005dd}. The quantity mainly investigated to this end is the two-particle source function, sometimes called the spatial correlation function (CF) or the pair-source distribution. Even though this quantity is not easy to reconstruct experimentally, detailed studies of its behaviour are merited by a multitude of reasons; including its connections to hydrodynamic expansion~\cite{Makhlin:1987gm,Cs\"org\H{o}:1995bi}, critical phenomena~\cite{Cs\"org\H{o}:2005it}, light nuclei formation~\cite{Oliinychenko:2020ply}, \textit{etc}. Phenomenological studies and experimental analyses, both, emphasise the importance of describing the shape of the source function. 
%\textcolor{blue}{$^{\rm Referee~3}$
Earlier, hydrodynamical, studies undertaken seemed to suggest a Gaussian source-shape~\cite{Cs\"org\H{o}:1995bi,Csan\'ad:2009wc}.
%}
Multiple measurements were also conducted with the Gaussian assumption~\cite{PhysRevC.71.044906,PHENIX:2004yan}. But, recent source-imaging studies suggest that the two-particle source function for pions has a long-range component, obeying a power-law behaviour~\cite{PHENIX:2017ino,PHENIX:2006nml,PHENIX:2007grx,PhysRevC.97.064912,L3:2011kzb,Kincses:2022eqq}.

%\textcolor{blue}{$^{\rm Referee~3}$
Femtoscopy~\cite{Lednicky2020} is the sub-field of high-energy heavy-ions physics that allows the investigation of the space-time structure of femtometre-scale processes encountered in high-energy nuclear- and particle-physics experiments.
%} 
%\textcolor{blue}{$^{\rm Referee~3}$
Femtoscopic correlations in heavy-ion collisions are currently understood to be caused partly by Bose-Einstein statistics~\cite{Bose1924,Einstein1925,PhysRevLett.3.181,PhysRev.120.300}. Alternatively, they are called Hanbury-Brown-Twiss (HBT) correlations in recognition of pioneering works by Hanbury-Brown and Twiss~\cite{doi:10.1080/14786440708520475,BROWN1956,HANBURYBROWN1956} on intensity interferometry in the field of observational astronomy to extract the apparent angular sizes of stars from correlations between the signals of two detectors.
%}
Additionally, correlations can arise out of final-state interactions, like electromagnetic interactions and strong-interactions, undergone by the investigated particles. These correlations between pairs of identical bosons can, hence, be used to explore the properties of the matter created in heavy-ion collisions and to map the geometry of the particle-emitting source~\cite{Lisa:2005dd}. 
%}

%%%%%%%%%%%%%%%%%%%%%%%%%%%%%%%%%%%%%%%%%%
\section{Correlations}
\label{s:femto}

Femtoscopy works on the principle that the momentum-correlation function of a pair of particles is related to the probability density of particle creation at a space-time point $X$, for a particle with four-momentum $P$. This probability density, $S(X, P)$, is also called the source-function. Defining $N_1(P)$ --- given by multiplying the particle-creation probability density with $\langle n \rangle$, the average number of particles --- as the invariant-momentum distribution and $N_2(P_1, P_2)$ --- given by multiplying the pair-creation probability density with $\langle n(n-1) \rangle$, the average number of pairs --- as the pair-momentum distribution, the two-particle correlation function can be written as~\cite{YANO1978556}:
%\begin{linenomath}
\begin{equation}
%    C(q) = 1 + \frac{\left|\Tilde{S}(q)\right|^2}{\left|\Tilde{S}(0)\right|^2}~;
    C(P_1, P_2) = \frac{N_2(P_1, P_2)}{N_1(P_1)N_1(P_2)}~;
\end{equation}
%\end{linenomath}
where
%\begin{linenomath}
\begin{equation}
    N_2(P_1, P_2) = \int S(X_1, P_1) S(X_2, P_2) \left| \psi_{P_1, P_2}(X_1, X_2) \right|^2 d^4 X_1 d^4 X_2~,
\end{equation}
%\end{linenomath}
with $\psi_{P_1, P_2}(X_1, X_2)$ being the two-particle wave-function that simplifies to:
%\begin{linenomath}
\begin{equation}
    \left| \psi_{P_1, P_2}(X_1, X_2) \right|^2 = \left| \psi^{(0)}_{P_1, P_2}(X_1, X_2) \right|^2 = 1 + \cos \left[ (P_1 - P_2) (X_1 - X_2) \right]
\end{equation}
%\end{linenomath}
in the interaction-free case for bosons, when only the quantum-statistical effects are taken into account. Thus, the correlation function can be re-defined as \cite{Csan\'ad:2019lkp}:
%\begin{linenomath}
\begin{equation}
    C(Q, K) \simeq 1 + \frac{\left|\Tilde{S}(Q,K)\right|^2}{\left|\Tilde{S}(0,K)\right|^2}~;
\label{eqn:C_q_S}
\end{equation}
%\end{linenomath}
where 
%\begin{linenomath}
\begin{equation}
    Q = P_1 - P_2~,~ K = \frac{P_1 + P_2}{2} 
\end{equation}
%\end{linenomath}
and
%\begin{linenomath}
\begin{equation}
    \Tilde{S}(Q,K) = \int S(X,K) e^{iQX} d^4 X
    \label{eqn:s_tilde}
\end{equation}
%\end{linenomath}
denote the pair-momentum difference, the average momentum and the Fourier-transform (FT) of the source, respectively, assuming that $Q \ll K$ holds for the kinematic range under investigation. The measurement of the correlation functions is done with respect to $Q$; over a range of well-defined $K$-values; and then the properties of the correlation functions are analysed as functions of the average-$K$, for each of those ranges.

%\noindent
%{\color{blue}
A significant fraction of the particles created in a heavy-ion collision is secondary, \textit{i.e.} they come from decays. Whereas the primordial particles are created directly from the hydrodynamic expansion of the collision-volume, the decay-particles arise from interactions that take place much later. Hence, the source can be assumed to consist of the following two components~\cite{Cs\"org\H{o}:1994in}:
\begin{enumerate}
    \item 
    a core, $S_{\rm C} (X,K)$ --- consisting of primordial particles created by the hydrodynamically-expanding, strongly-interacting quark-gluon plasma; along with the decays of resonances with half-lives less than a few fm/$c$; and
    \item 
    a halo, $S_{\rm H} (X,K)$ --- consisting of the products created by the decay of long-lived resonances, including but not limited to $\eta$, $\eta'$, $K_S^0$, $\omega$,
\end{enumerate}
making it possible to decompose $S (X,K)$ as~\cite{Cs\"org\H{o}:1994in}:
%}
%\begin{linenomath}
\begin{equation}
    S (X,K) = S_{\rm C} (X,K) + S_{\rm H} (X,K)~.
\label{eqn:core-halo_S}
\end{equation}
%\end{linenomath}
%\textcolor{green}{$^{\rm Referee~1}$
As explained in detail in Ref.~\cite{Cs\"org\H{o}:1994in}, the Fourier-transformed emission function of the halo vanishes for resolvable relative momenta, \textit{i.e.} the $Q$-values that lie in the experimentally achievable region. Hence, the halo does not contribute to the measured correlation function, which in turn is determined by the core component. Hence, the measured correlation function, when extrapolated to $Q=0$, does not take a value of 2 as expected from Eqn. (\ref{eqn:C_q_S}), but takes the form:
%\begin{linenomath}
\begin{equation}
    \lim_{Q\to0} C(Q,K) = 1 + \lambda (K) < 2~.
\label{eqn:C_q_lam}
\end{equation}
%\end{linenomath}
This ``intercept parameter" of the correlation function, also called the correlation-strength, $\lambda(K)$, may depend on the pair-momentum $K$ and can be understood on the basis of the core-halo model by re-writing the correlation function using Eqns. (\ref{eqn:C_q_S}), (\ref{eqn:s_tilde}) and (\ref{eqn:core-halo_S}):
%}
%\begin{linenomath}
\begin{align}
    C(Q,K) &= 1 + \left[ \frac{N_{\rm C}(K)}{N_{\rm C}(K) + N_{\rm H}(K)} \right]^2 \frac{\left|\Tilde{S}_{\rm C}(Q,K)\right|^2}{\left|\Tilde{S}_{\rm C}(0,K)\right|^2} \nonumber \\
                &= 1 + \lambda (K) \frac{\left|\Tilde{S}_{\rm C}(Q,K)\right|^2}{\left|\Tilde{S}_{\rm C}(0,K)\right|^2}~,
\end{align}
%\end{linenomath}
with $N_{\rm C}(K) = \int S_{\rm C}(X,K) d^4 X$ and $N_{\rm H}(K)= \int S_{\rm H}(X,K) d^4 X $ being, respectively, the core's and the halo's contributions and with $\lambda(K)$ being:
%\begin{linenomath}
\begin{equation}
    \lambda (K) = \left[ \frac{N_{\rm C}(K)}{N_{\rm C}(K) + N_{\rm H}(K)} \right]^2~.
\label{eqn:lam}
\end{equation}
%\end{linenomath}
Realising that $X \equiv X(\vec{r},t)$, the spatial correlation function:
%\begin{linenomath}
\begin{equation}
    D(\vec{r},K)= \int S(\vec{r'}+ \frac{\vec{r}}{2},K)S(\vec{r'}- \frac{\vec{r}}{2},K)d^3 \vec{r'}
\end{equation}
%\end{linenomath}
can be  used to re-write $C(Q,K)$ as \cite{Csan\'ad:2019lkp}:
%\begin{linenomath}
\begin{equation}
    C(Q,K) \approx \frac{\int D(\vec{r},K) \left| \psi_{Q}(\vec{r}) \right|^2 d^3 \vec{r}}{\int D(\vec{r},K) d^3 \vec{r}}~.
\end{equation}
%\end{linenomath}

%%%%%%%%%%%%%%%%%%%%%%%%%%%%%%%%%%%%%%%%%%
\section{L\'evy-distribution}
\label{s:levy}

Usually, the shape of the source distribution is assumed to be Gaussian. However, evidence of a non-Gaussian source for correlated pions has been found in multiple studies, necessitating a generalisation of the distribution to a L\'evy-stable one \cite{NOLAN1998187}:

%\begin{linenomath}
\begin{equation}
    S(x,K) = \mathcal{L} \left( x;\alpha, \lambda, R \right) = \frac{1}{2\pi}\int e^{-(\mathcal{Q'}R)^\alpha} e^{i\mathcal{Q'}x} d\mathcal{Q'}~;
\label{eqn:levy_distro_main}
\end{equation}    
%\end{linenomath}
where $\mathcal{L}$ is the one-dimensional L\'evy-function, $R$ is the L\'evy-scale-parameter, $\lambda$ is the correlation-strength, $0<\alpha\leq 2$ is the L\'evy-exponent and $\mathcal{Q'}$ is the integration-variable. These parameters are, generally, understood to depend on $K$. Salient features of the distribution include moments greater than $\alpha$ being undefined and $D (\vec{r},K)$ necessarily having a L\'evy-shape, with the same $\alpha$, in cases where $S (X,K)$ is L\'evy-shaped. The distribution exhibits a power-law behaviour for $\alpha < 2$, with $\alpha = 1$ representing a Cauchy distribution and $\alpha = 2$ representing a Gaussian distribution. Multiple factors; such as anomalous diffusion, jet-fragmentation and proximity to the critical end point (CEP); can contribute to the appearance of L\'evy-stable sources. However, the high-multiplicity, nucleon-on-nucleon nature of the heavy-ion collisions analysed makes it unlikely for jet-fragmentation to be the dominant reason for the appearance of L\'evy-sources in this study --- as it has been identified as the cause behind L\'evy-stable sources in $e^+e^-$ collisions at LEP \cite{L3:2011kzb}. On the other hand, the high centre-of-mass-energy of the collisions explored here rules out the possibility of the system being close to the critical end point~\cite{Cs\"org\H{o}:2004sr,Cs\"org\H{o}:2005it,Csan\'ad:2007fr}. 

Interestingly, it is trivial to establish that $\alpha$ is related to one of the critical exponents, $\eta$. In case of a second-order phase transition, the $\eta$ exponent describes the power-law behaviour of the spatial correlation function at the critical end point with an exponent of $-(d - 2 + \eta)$; where $d$ is the dimension. In a three-dimensional analysis, with $d=3$, this exponent would compute to $-(1 + \eta)$. However, it is established that the three-dimensional L\'evy-distribution describes the power-law tail of the spatial correlation function with an exponent of $-(1 + \alpha)$. Thus, comparing the exponents at the critical end point, it can be easily seen that the L\'evy-exponent, $\alpha$, is identical to the critical exponent, $\eta$; a conjecture explored in Ref.~\cite{Cs\"org\H{o}:2008ayr}. The second-order QCD phase transition is expected to be in the same universality class as the three-dimensional Ising model and in that case, the $\eta$-exponent has a  value of 0.03631$\pm$0.00003 at the critical end point~\cite{El-Showk:2014dwa}. However, the universality class of the random-field, three-dimensional Ising model may also be of relevance here, where the value of $\eta$ is 0.50$\pm$0.05~\cite{Rieger1995}. Thus, extracting $\alpha$ at collision energies lower than the ones used in this analysis; while taking into account finite-size and finite-time effects; might yield insightful information about the nature of the quark-hadron phase transition and shed light on the location of the critical end point in the QCD phase diagram~\cite{PhysRevD.58.096007,PhysRevLett.81.4816,Cs\"org\H{o}:2005it,Lacey:2015yxg,lacey2016finitesize}.

As coordinate-space distributions extracted from experimental data show a heavy tail, the limitations of the hydrodynamical approach --- assuming idealised freeze-out with a sudden jump in the mean-free-path from zero to infinity --- become clear. This requires the, more realistic, approach using hadronic re-scattering where the system cools as it dilutes with an expanding hadron gas, its mean-free-path diverges to infinity in a finite time-interval and re-scattering occurs in the presence of a time-dependent mean-free-path. This signals the existence of anomalous diffusion --- experimentally observed as power-law-shaped tails in coordinate-space distributions of the source --- in the system; as opposed to normal diffusion --- with the Gaussian source exhibiting a strongly-decaying tail --- caused by the Brownian motion of the particles constituting the system.

One-dimensional, normal diffusion's momentum-space diffusion-equation is \cite{Csan\'ad:2007fr}:
%\begin{linenomath}
\begin{equation}
    \frac{\partial W(q,t)}{\partial t} = -k_{\rm N} q^2 W(q,t)~;%\nonumber
\label{eqn:norm_diff_mom}    
\end{equation}
%\end{linenomath}
where $k_{\rm N}$ is the normal-diffusion constant, $q$ is the momentum, $t$ is the time and $W(q,t)$ is the momentum-space probability distribution. The coordinate-space solution to Eqn. (\ref{eqn:norm_diff_mom}) is given by the Gaussian expression: 
%\begin{linenomath}
\begin{equation}
    W\left( x,t \right) = \frac{1}{\sqrt{4 \pi k_{\rm N} t}} \exp{\left( -\frac{x^2}{4 k_{\rm N} t} \right)}~.
\end{equation}    
%\end{linenomath}       
For anomalous diffusion, the coordinate-space diffusion-equation; in terms of the spatial probability distribution $W\left( x,v,t \right)$; is the generalised Fokker-Planck equation \cite{Csan\'ad:2007fr}:
%\begin{linenomath}
\begin{equation}
     \frac{\partial W}{\partial t} + v \frac{\partial W}{\partial x} + \frac{F\left( x \right)}{m} \frac{\partial W}{\partial v} = \eta_{\alpha'0} D_t^{1 - \alpha'} L_{\rm FP} W~.
     \label{eqn:FP}
\end{equation}
%\end{linenomath}
%\textcolor{green}{$^{\rm Referee~1}$}\textcolor{blue}{$^{\rm Referee~3}$
Here, $\eta_{\alpha'}$ is the generalised friction constant of dimension $\left[\eta_{\alpha'}\right] = s^{\alpha' - 2}$,\\ $\prescript{}{0}{D_t^{1 - \alpha'}}$ is the Riemann-Liouville operator:
%\begin{linenomath}
\begin{equation}
    \prescript{}{0}{D_t^{1 - \alpha'}} t^p = \left(\frac{\partial}{\partial t}\right) \prescript{}{0}{D_t^{- \alpha'}} t^p = \frac{\Gamma(1+p)}{\Gamma(p+\alpha')} t^{p+\alpha'-1}
\end{equation}
%\end{linenomath}
and $L_{\rm FP}$ is the Fokker-Planck operator:
%\begin{linenomath}
\begin{equation}
    L_{\rm FP} = \frac{\partial}{\partial x} \frac{V'(x)}{m \eta_{\alpha'}} + k_{\rm A} \frac{\partial^2}{\partial x^2}~,
\end{equation}
%\end{linenomath}
with $V'(x)$ being related to the force $F(x)$ by $F(x) = -\frac{dV(x)}{dx}$; as explained in Refs.~\cite{METZLER20001,Cs\"org\H{o}:2003uv}.
%}
The momentum-space solution to Eqn. (\ref{eqn:FP}) is given by: 
%\begin{linenomath}
\begin{equation}
    W(q,t) = \exp{\left( -t k_{\rm A}^\alpha |q|^\alpha \right)}~.
\end{equation}
%\end{linenomath}
This $W(q,t)$ happens to be the FT; or the characteristic function; of L\'evy-stable source-distributions, with $\alpha$ being the L\'evy-exponent from Eqn. (\ref{eqn:levy_distro_main}) and $k_{\rm A}$ being the anomalous-diffusion constant. If a centred, spherically symmetric, L\'evy-stable source-distribution is assumed; and all final-state interactions are neglected; the one-dimensional, two-particle correlation function takes the simplified form:
%\begin{linenomath}
\begin{equation}
    C(q) = 1 + \lambda \cdot e^{-(qR)^\alpha}~,
\label{eqn:C_q_fit}    
\end{equation}
%\end{linenomath}
with $\lambda$ as the correlation-strength from Eqns.~(\ref{eqn:lam}) and (\ref{eqn:C_q_lam}), $R$ as the L\'evy-scale, $\alpha$ as the L\'evy-exponent and $q$ as the absolute value of the three-momentum-difference in the longitudinally co-moving system (LCMS)~\cite{PHENIX:2017ino}:
%\begin{linenomath}
\begin{equation}
    q = {q}_{\rm LCMS} = |\vec{p}_1-\vec{p}_2|_{\rm LCMS}~.
\end{equation}
%\end{linenomath}
%\textcolor{blue}{$^{\rm Referee~3}$
$R$ can be interpreted as the homogeneity-length of the particle-species, while $\alpha$ represents the extent of the anomalous diffusion occurring in the system.
%}
The spherical symmetry in ${q}_{\rm LCMS}$ is ideal for a one-dimensional analysis of a three-dimensional, spherically symmetric system. Subsequent measurements, with higher precision, might necessitate a move towards a full, three-dimensional analysis. Until then, the approximations used in Ref.~\cite{kurgyis2023coulomb} may be utilised for a preliminary analysis.

%\textcolor{green}{$^{\rm Referee~1}$
Momentum-correlations of like-sign kaon-pairs, at $\sqrt{s_{_{\rm NN}}} = 200$ GeV, can be utilised to calculate $C(q)$ and to, thus, ascertain the shape of the pair-source distribution. If anomalous diffusion is the sole source of non-Gaussianity, then it is expected that the extent of the anomaly will depend on the total cross-section, or equivalently, on the mean-free-path. Since the mean-free-path is larger for kaons than for pions, the former's diffusion is expected to be more anomalous in the hadron gas. Hence, the $\alpha$ for kaons, $\alpha_\kappa$, is expected to be smaller~\cite{Csan\'ad:2007fr}. Thus, measuring $\alpha_\kappa$ may shed light on the role of anomalous diffusion in the hadron gas as the origin of the appearance of L\'evy-distributions.
%}

%%%%%%%%%%%%%%%%%%%%%%%%%%%%%%%%%%%%%%%%%%
\section{Measurement}
\label{s:measure}

\begin{figure}
\centering
\begin{subfigure}{.5\textwidth}
  \centering
  \includegraphics[width=1\textwidth]{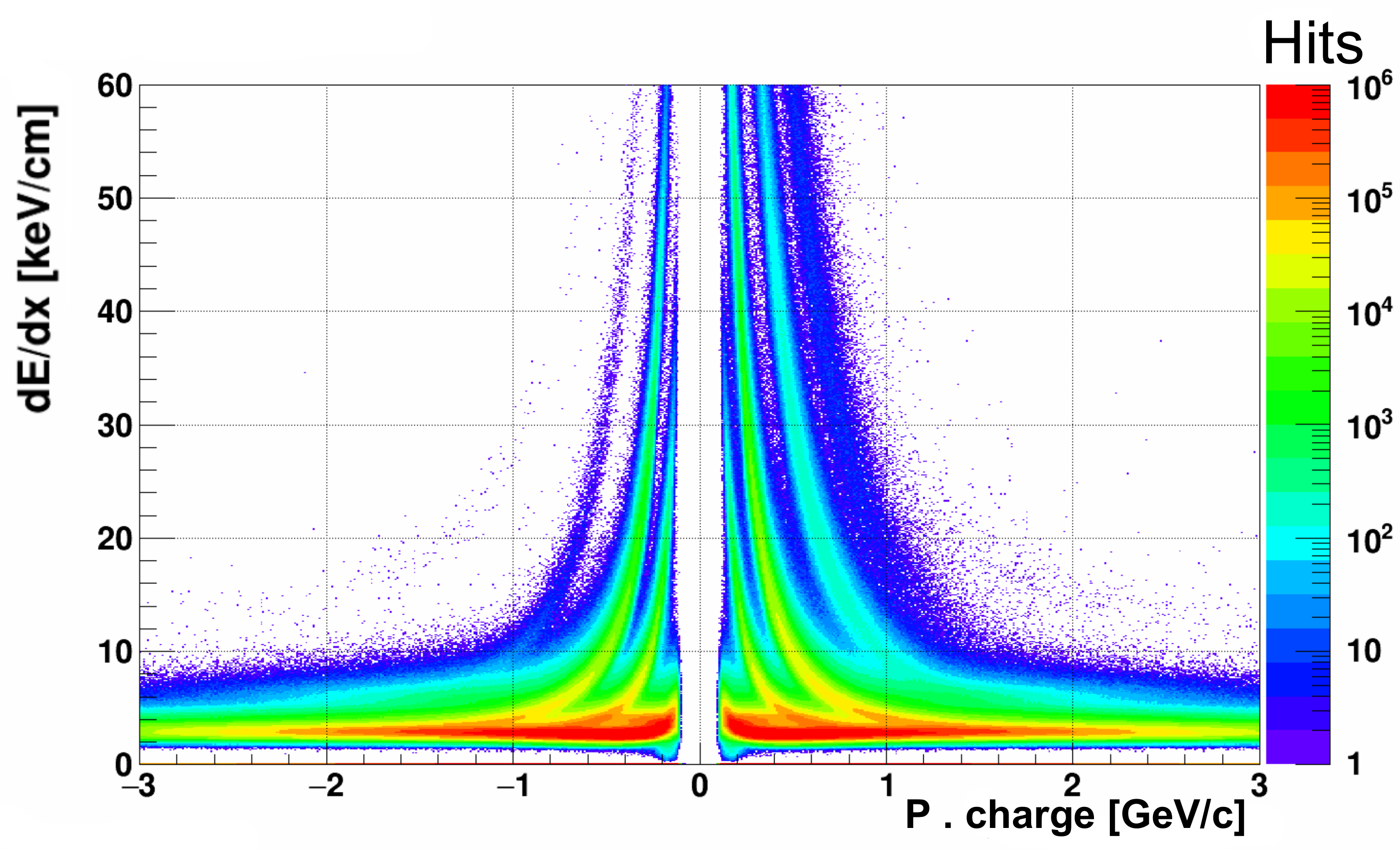}
  \label{fig:dedx_bef}
  \caption{}
\end{subfigure}%
\begin{subfigure}{.5\textwidth}
  \centering
  \includegraphics[width=1\textwidth]{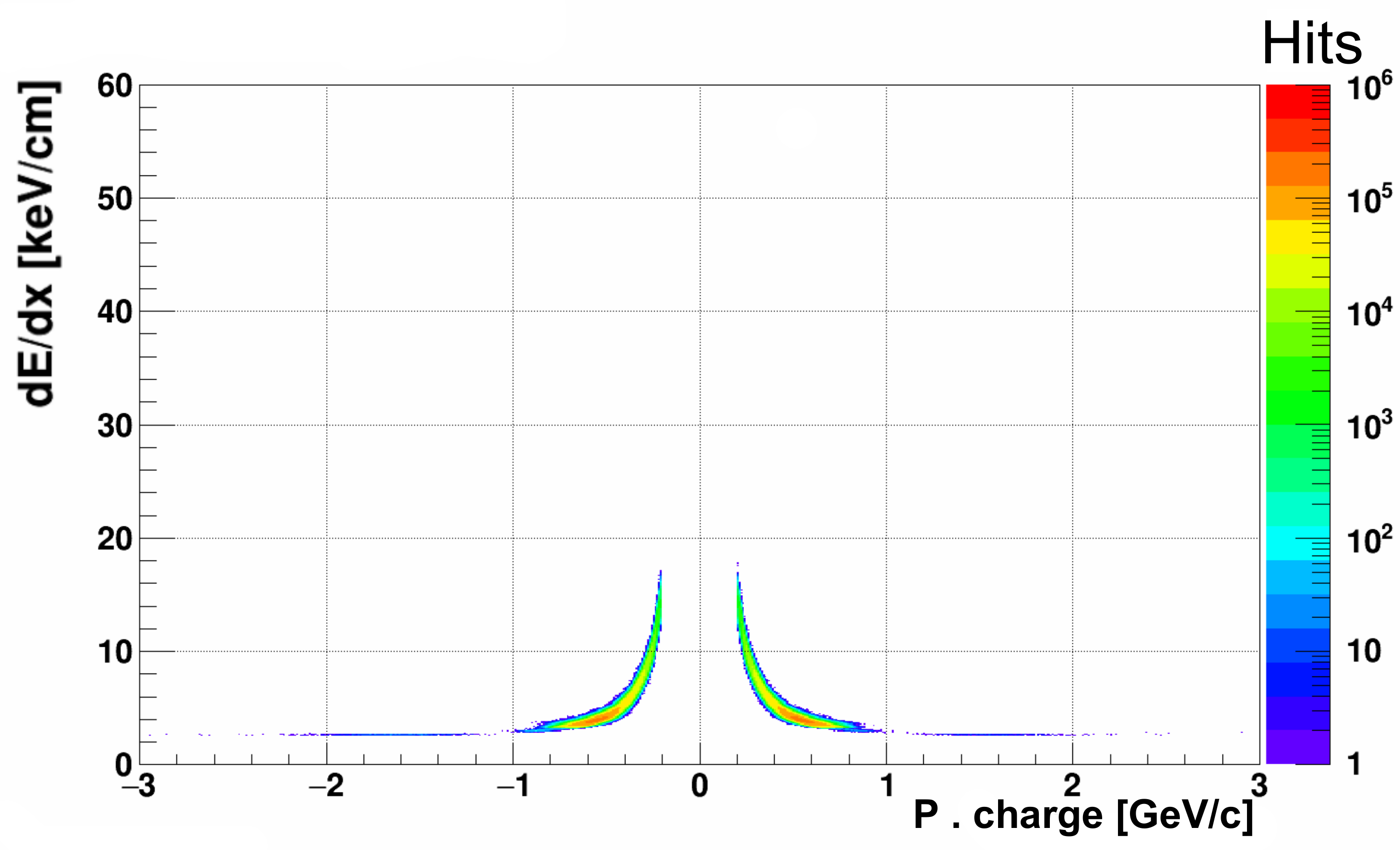}
  \label{fig:dedx_aft}
  \caption{}
\end{subfigure}
\caption{
%\textcolor{blue}{$^{\rm Referee~3}$
Sample ionisation-energy-loss, as a function of momentum$\times$charge, \textbf{(a)} for all the charged particles available and \textbf{(b)} after being cut only for the charged kaons to be isolated.
%}
}
\label{fig:dedx}
\end{figure}

The data used for this analysis is obtained from RHIC's gold-on-gold collisions at 200 GeV centre-of-mass energy per nucleon pair; performed in 2016; as measured by the solenoidal tracker at RHIC (STAR) experiment. Different particle species, depending on their mass and charge, produce different shapes when their ionisation-energy-loss,  $dE/dx$, is plotted as a function of momentum. These shapes can help distinguish the particle species from each other and help isolate the kaons, as observed in Fig. \ref{fig:dedx}. For this investigation, the analysis processes about 410 million events in the 0-30\% centrality range. They undergo strict track- and pair-selection criteria that, following Ref.~\cite{PhysRevC.92.014904}, including the identification of kaons via energy-loss measured by the time-projection chamber (TPC); pair-selection based on the fraction-of-merged-hits (FMH) and the splitting-level (SL); and limitations on the track's momentum, rapidity and distance-of-closest-approach (DCA).

%\noindent
The one-dimensional, like-sign, two-kaon, femtoscopic correlation functions are then experimentally constructed, using the event-mixing technique \cite{Kincses:2019czd}.
The actual pair-distribution, $A(q)$, is formed from kaon-pairs, with members of the pair belonging to the same event. This distribution is affected by various effects, such as kinematics and acceptance. To correct for these effects, a background-distribution, $B(q)$, is constructed with the pairs, where the members originate from separate events. In this analysis, the method for event-mixing described in Refs.~\cite{PHENIX:2017ino,L3:2011kzb} is used to correct for residual correlations. For a set of event-classes based on the centrality and the location of the collision-vertex, a background event pool is established. Then, for each real event, a mixed event of the same event-class is created from this pool, making sure that each particle in this mixed event belongs to a different real event. Subsequently, pairs within the mixed event contribute to the formation the aforementioned background-distribution, $B(q)$. Finally, the pre-normalised correlation function is calculated as:
    %\begin{linenomath}
    \begin{equation}
        C(q) = \frac{A(q)}{B(q)} \cdot \frac{\int B(q) \ dq}{\int A(q) \ dq}~,
    \end{equation}
    %\end{linenomath}
for three different ranges of transverse-mass, $m_{\rm T}$, defined as $m_{\rm T} = \sqrt{m^2 + (K_{\rm T}/c)^2}$; with $m$ as the kaon-mass and $K_{\rm T}$ as the average transverse-momentum of the pair. The normalisation integrals are performed over a range where the correlation function is not supposed to exhibit quantum-statistical features. It is to be noted that, the method outlined here is applied to pairs belonging to a given range of average momenta. Furthermore, in the event-mixing technique described above,
the number of actual and background pairs is the same,
aside from the effect of the pair-selection criteria mentioned earlier.

%\noindent
With the momentum correlations, obtained from experimental data, measured and the empirical values of the correlation function calculated; preparations are made for fitting the L\'evy-function, detailed in Eqn. (\ref{eqn:C_q_fit}), to $C(q)$.
%so calculated. 
However, the assumptions behind Eqn. (\ref{eqn:C_q_fit}) make it unsuitable for a direct fit to experimentally-obtained data. As mentioned in Sec. \ref{s:intro}, final-state interactions have considerable effects on the momentum correlations between like-sign kaon-pairs and hence, they need to be imbibed into the analysis as corrections to Eqn. (\ref{eqn:C_q_fit}) --- to make it suitable and, physically, relevant as a fit for the correlation function obtained above. Multiple factors can contribute to the final-state modifications of the momentum-correlations, but leading amongst them are Coulomb interactions; since a gas of charged hadrons can never be, entirely, devoid of Coulomb repulsion.

\begin{figure}[!t]
    \includegraphics[width=0.8\textwidth]{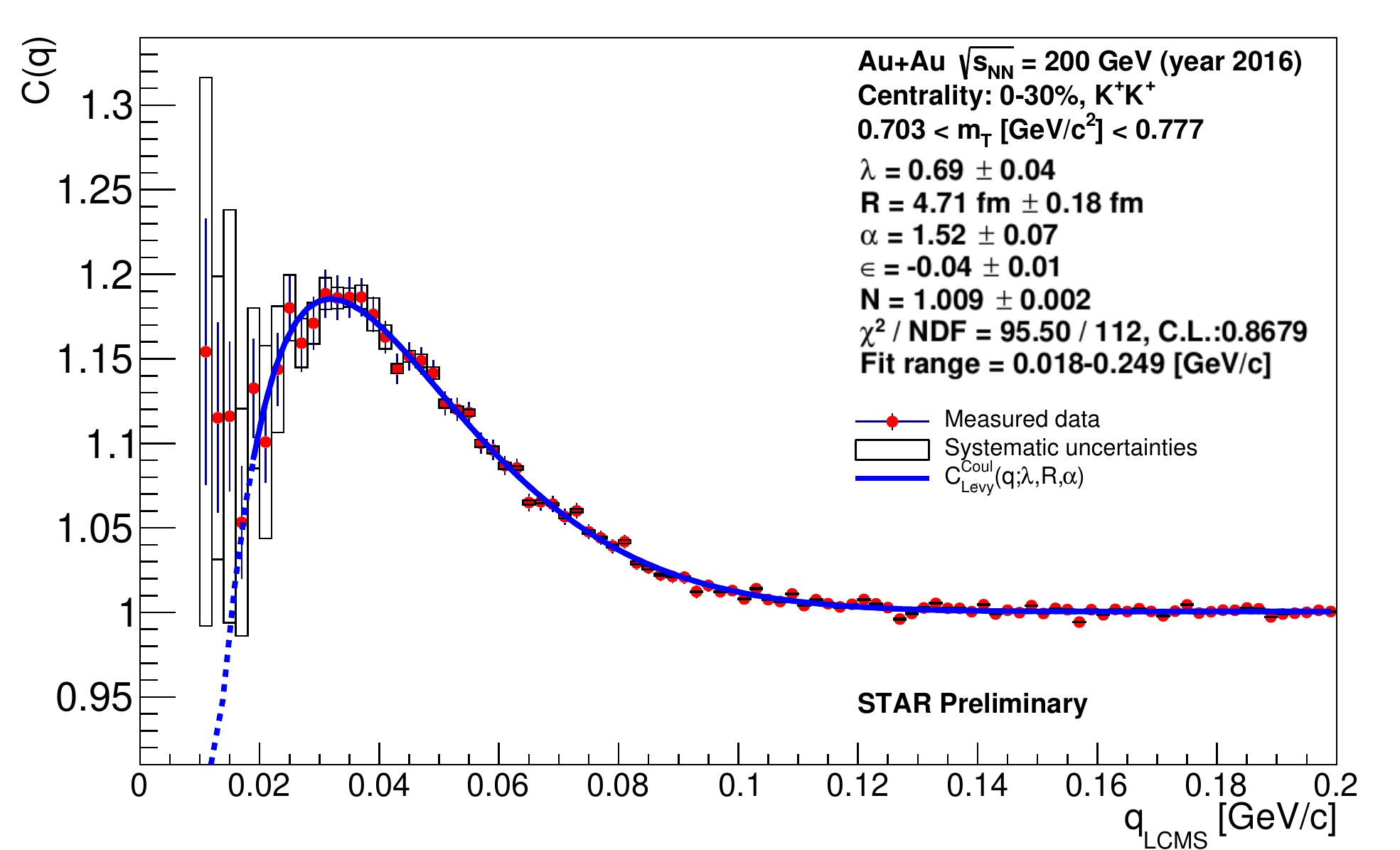}
    \caption{$C(q)$, as a function of $q_{\rm LCMS}$, for positively-charged kaon-pairs in the $m_{\rm T}$ range: 703--777 MeV/$c^2$ and the centrality range: 0--30\%. The red dots denote the measured data and the blue lines (solid \& dotted) denote the fit. The systematic uncertainties are shown as hollow rectangles.}
    \label{fig:C_q_bin3}
\end{figure}   
%\unskip

\begin{figure}
    \includegraphics[width=0.8\textwidth]{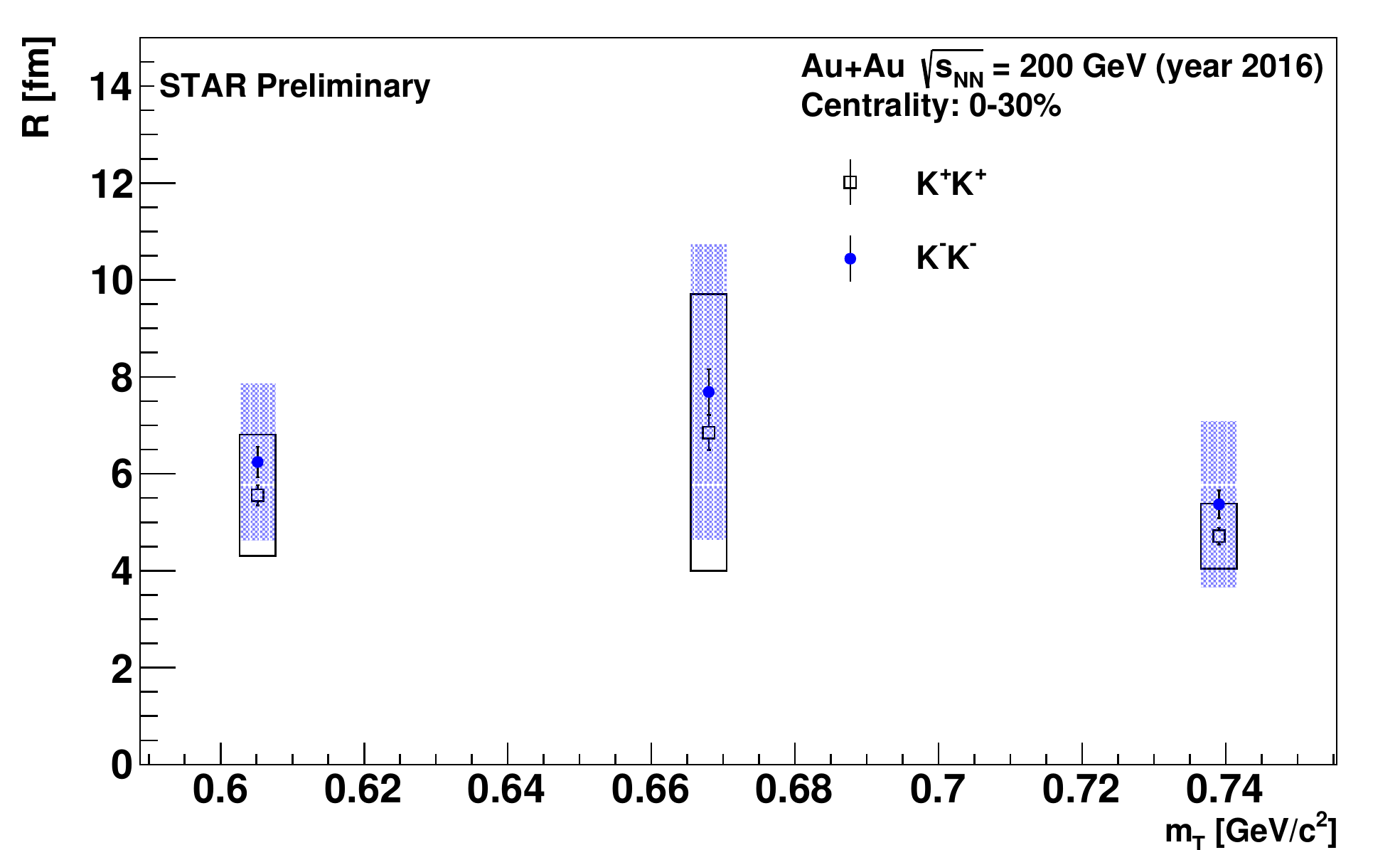}
    \caption{$R$, as a function of $m_{\rm T}$, for 0--30\% centrality. The hollow, blue squares denote positively-charged kaon-pairs and the solid, blue circles denote negatively-charged kaon-pairs; along with their error-bars. The systematic uncertainties are shown as hollow ($K^+K^+$) and shaded ($K^-K^-$) rectangles.}
    \label{fig:rad}
\end{figure}
%\unskip

%\noindent
The final-state Coulomb-interactions are incorporated into the CF by using the Bowler-Sinyukov formula, that includes a correction-term for Coulomb-repulsion, as \cite{BOWLER199169,Sinyukov:1998fc}:
%\begin{linenomath}
\begin{equation}
    C(q) = \left[1 - \lambda + \lambda \cdot \mathcal{K}(q; \alpha, R) \cdot \left(1 + e^{-(qR)^\alpha} \right) \right] \cdot N \cdot (1 + q \varepsilon)~,
\end{equation}
%\end{linenomath}
where $N$ is a normalisation factor, $\varepsilon$ is responsible for a linear, residual, long-range background and $\mathcal{K}$ is the Coulomb correction \cite{Csan\'ad:2019lkp}:
%\begin{linenomath}
\begin{equation}
    \mathcal{K}(q; \alpha, R) = \frac{\int D(\vec{r}) \left| \psi_{q}^{\rm Coul}(\vec{r}) \right|^2 d^3 \vec{r}}{\int D(\vec{r}) \left| \psi_{q}(\vec{r}) \right|^2 d^3 \vec{r}}~,
\end{equation}
%\end{linenomath}
with $D(\vec{r})$ being the, simplified, spatial pair-distribution and $\psi_{q}^{\rm Coul}(\vec{r})$ being the solution to the two-particle Schr\"odinger-equation, in the presence of a Coulomb potential. In this study, $\mathcal{K}(q; \alpha, R)$ for kaons is calculated numerically employing the procedure used in Refs.~\cite{Csan\'ad:2019lkp,PhysRevC.97.064912,Csan\'ad:2019cns}.
The inclusion of other final-state contributions, like the strong interaction, can resolve the possible underestimation --- regarding $R$ and $\lambda$ --- and overestimation --- regarding $\alpha$ --- of the L\'evy-parameters \cite{Kincses:2019rug}; but, the statistical significance of such precise corrections turns out to be negligible in the context of the current measurement.

\begin{figure}%[!b]
    \includegraphics[width=0.8\textwidth]{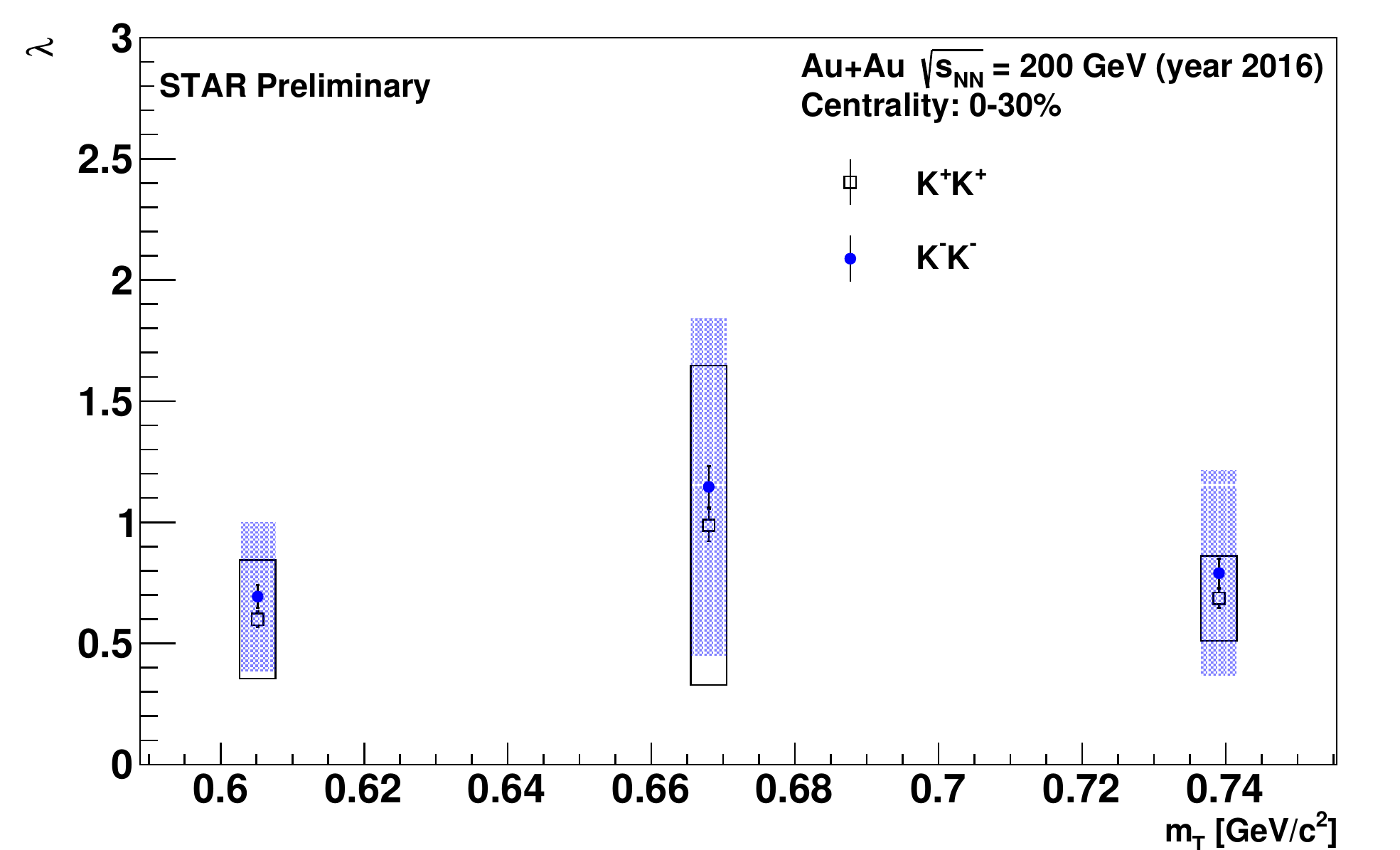}
    \caption{$\lambda$, as a function of $m_{\rm T}$, for 0--30\% centrality. The hollow, blue squares denote positively-charged kaon-pairs and the solid, blue circles denote negatively-charged kaon-pairs; along with their error-bars. The systematic uncertainties are shown as hollow ($K^+K^+$) and shaded ($K^-K^-$) rectangles.}
    \label{fig:lam}
\end{figure}
%\unskip

%%%%%%%%%%%%%%%%%%%%%%%%%%%%%%%%%%%%%%%%%%
\section{Results \& Discussion}
\label{s:res}

As illustrated by Fig. \ref{fig:C_q_bin3}, the Coulomb-corrected L\'evy-distribution function agrees with the measured $C(q)$ over the entire $q_{\rm LCMS}$-range. The femtoscopic peak \cite{PhysRevLett.3.181,PhysRev.120.300} and the Coulomb-hole \cite{Sinyukov:1998fc} are both observed, as expected. 
%The peak is seen to be increasing with an increase in the transverse-mass. 
%The up-tick in $C(q)$ at small-$q_{\rm LCMS}$ is a numerical artefact, which is in the process of being removed. 
The values of the normalisation factor, $N$, and the linear-background factor, $\varepsilon$, are observed to be close to 1 and 0, respectively.
%tentatively disproving the assumption of the linear background.

%\noindent
The systematic uncertainties are obtained by combining the uncertainties arising from variations in the event- and pair-selection criteria, denoted by $\Delta_{\rm cuts}$, mentioned above and those arising out of variations to the range of the fit, denoted by $\Delta_{\rm fits}$. At this preliminary stage, systematic uncertainties arising out of variations in the track-selection criteria are not included. Thus, the final systematic-uncertainties, $\Delta_{\rm total}$, are obtained as:
%\begin{linenomath}
\begin{equation}
    \Delta_{\rm total} = \sqrt{(\Delta_{\rm fits})^2 + (\Delta_{\rm cuts})^2}~.
\end{equation}
%\end{linenomath}

\begin{figure}%[H]
    \includegraphics[width=0.8\textwidth]{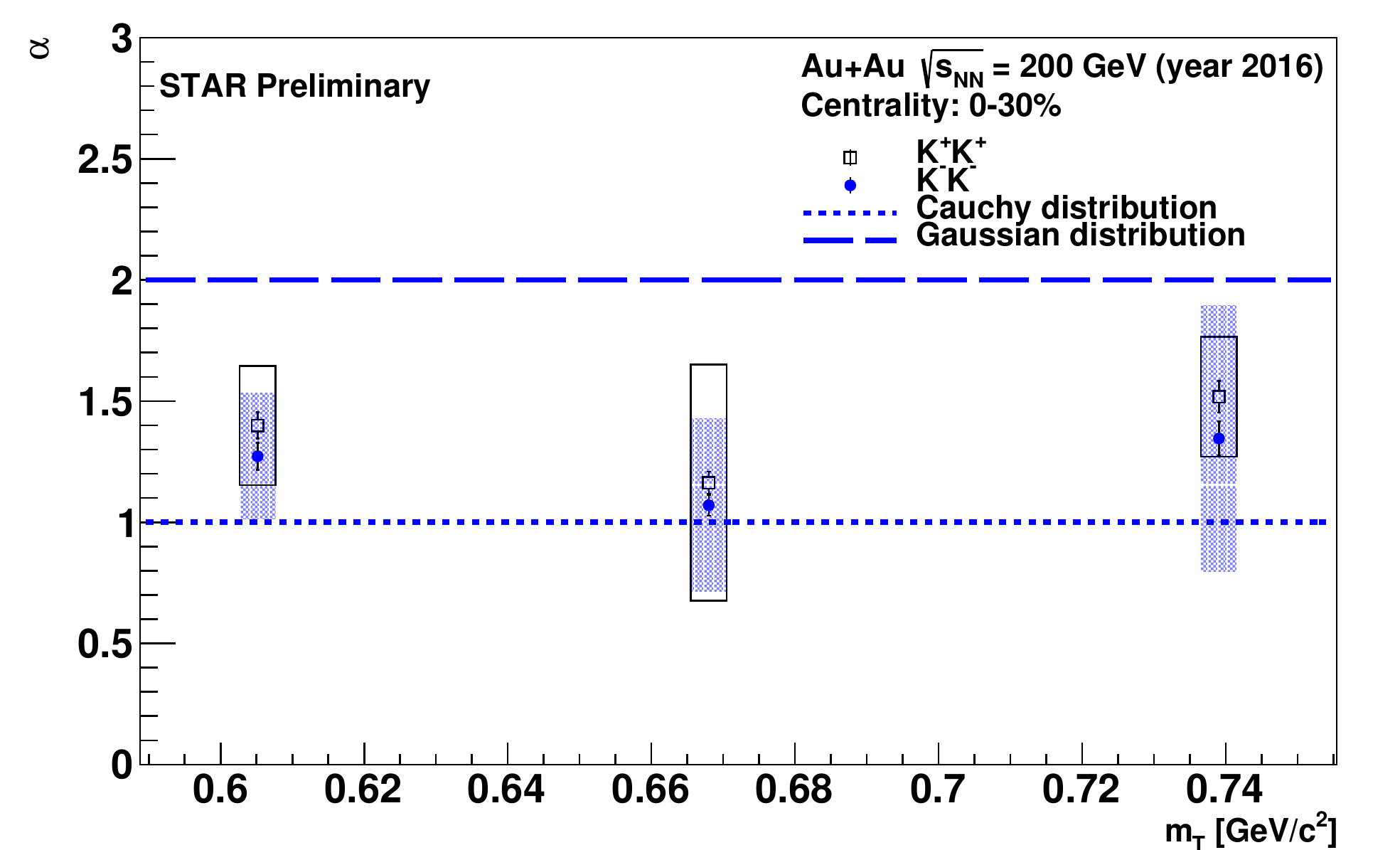}
    \caption{$\alpha$, as a function of $m_{\rm T}$, for 0--30\% centrality. The hollow, blue squares denote positively-charged kaon-pairs and the solid, blue circles denote negatively-charged kaon-pairs; along with their error bars. The systematic uncertainties are shown as hollow ($K^+K^+$) and shaded ($K^-K^-$) rectangles.}
    \label{fig:alp}
\end{figure}
%\unskip

Fig. \ref{fig:rad} shows the kaon-homogeneity length, $R$; otherwise known as the L\'evy-scale; as a function of $m_{\rm T}$. It is observed to exhibit large, systematic uncertainties; a very weak dependence on $m_{\rm T}$; and a possible decrease w.r.t. it, as found from previous studies~\cite{PHENIX:2004yan,PHENIX:2009ilf,Cs\"org\H{o}:1995bi,Makhlin:1987gm,Csan\'ad:2009wc,Kincses:2022eqq}. 
%\textcolor{green}{$^{\rm Referee~1}$
Note, however, that the hydrodynamical studies predicting the decrease of the L\'evy-scale, as a function of $m_{\rm T}$, are based on the Gaussian-source assumption~\cite{Cs\"org\H{o}:1995bi,Makhlin:1987gm}.
%}
Hence, more investigations on this topic are in order. The extracted values of the L\'evy-scale, in this charged-kaon analysis, are also found to be similar to PHENIX's like-sign pion results~\cite{PHENIX:2017ino}, with $R_\pi \sim 5$--7 fm for the $m_{\rm T}$-range of 600--700 MeV/$c^2$. A more detailed comparison of the $m_{\rm T}$-dependence of L\'evy-scales of different particle-species could shed light on the origin of the appearance of L\'evy-stable sources, given that; from calculations based on hydrodynamics; a species-independent $m_{\rm T}$-scaling was predicted in Ref.~\cite{Csan\'ad:2008gt}.

%\noindent
The intercept of the correlation function --- the correlation strength, $\lambda$ --- is shown in Fig. \ref{fig:lam}. Values extracted from the fits show that it's close to unity; as is to be expected from the small fraction of decay-kaons present in the system.

%\noindent
The extent of the anomalous diffusion might be gleaned from the L\'evy-exponent, $\alpha$, as seen in Fig. \ref{fig:alp}. It also illustrates the values corresponding to the Gaussian and Cauchy distributions, with dashed and dotted blue lines, respectively. The L\'evy-exponent is observed to have values between those two extremes, thereby indicating a power-law behaviour and the, likely, existence of anomalous diffusion. The extracted values of $\alpha \sim$ 1.0--1.5 for kaons are, again, similar to PHENIX's pion-results; with $\alpha_\pi \sim 1.2$ in the same transverse-mass range. $\alpha_\kappa$ not being smaller than $\alpha_\pi$ hints at the existence of other factors, on top of anomalous diffusion, contributing to the appearance of non-Gaussian source-shapes. However, the current statistics prevent the drawing of more definitive conclusions.
%at this stage. 

%\newpage
%%%%%%%%%%%%%%%%%%%%%%%%%%%%%%%%%%%%%%%%%%
\section{Summary \& Outlook}
\label{s:sum}

Preliminary analysis of data collected by STAR, from RHIC's 2016 BES $\sqrt{s_{_{\rm NN}}} = 200$ GeV $\mathrm{Au}+\mathrm{Au}$ collisions, suggests a non-Gaussian, L\'evy-stable source-shape for pairs of the identically-charged kaons produced in the collisions. 
The L\'evy-stability-exponent $\alpha_\kappa$, is observed to be comparable to that of like-sign pion-pairs obtained from PHENIX. 

%\noindent
However, anomalous diffusion may not be solely responsible for the heavy tails observed in the source distributions, as suggested by $\alpha_\kappa$ being comparable to $\alpha_\pi$. It is to be noted that, a complete systematic uncertainty analysis; which is currently ongoing; is required to achieve definitive conclusions about any and all claims made. Seeing as L\'evy-stable sources can arise in strongly-interacting systems due to their proximity to the QCD critical end point, at higher chemical potentials, similar studies at lower beam energies would likely strengthen the search for the QCD critical end point.
%at this stage. 

%\newpage
%%%%%%%%%%%%%%%%%%%%%%%%%%%%%%%%%%%%%%%%%%
\section*{Acknowledgement}

\noindent
This project has been funded, in part, by the TKP2021-NKTA-64 and the K-136138 grants from the NKFIH, Hungary and, in part, by the Department of Energy of the United States of America. This paper was prepared for a Special Issue corresponding to the Zimányi School Winter Workshop 2022. The author would like to thank the STAR collaboration and the people behind the RHIC framework, whose incessant efforts have produced every bit of data analysed as a part of this project. The author is also, deeply, indebted to Koushik Mandal for his invaluable help with discussions about the ROOT framework and the C++ programming language; in particular; and about the myriad intricacies of navigating the field of experimental heavy-ions physics; in general.

\end{document}